\documentclass[11pt]{article} 
\usepackage{hyperref} 
\pdfoutput=1 
\usepackage{hyperref} 
\pdfoutput=1 

\usepackage{graphicx}

\begin{document} 
\title{Visualization of Kelvin waves on quantum vortices}

\author{Enrico Fonda$^{1,2,3}$, David P. Meichle$^{1}$, Nicholas T. Ouellette$^{4}$, \\
Sahand Hormoz$^{5}$, Katepalli R. Sreenivasan$^{3}$, and Daniel P. Lathrop$^{1}$\\
\\\vspace{6pt}
1) University of Maryland, College Park, USA,\\
2) Universit\`{a} degli studi di Trieste, Italy, \\
3) New York University, USA,\\
4) Yale University, USA,\\
5) University of California, Santa Barbara, USA,\\
}

\maketitle

\begin{abstract} 
In superfluid helium, vorticity is quantized and constrained on line-like phase singularities, called quantum or quantized vortices. By visualizing the motion of sub-micron frozen particles in superfluid $^{4}$He, we directly observe for the first time the helical Kelvin waves excited after quantized vortex reconnections. We compare the data with self-similar solutions of vortex filament models. We report the results in a fluid dynamic video.
\end{abstract} 

In superfluid helium, vorticity is quantized and constrained on line-like phase singularities, called quantum or quantized vortices. The evolution of a tangle of these topological defects is called `quantum turbulence' \cite{Donnelly1991}. 
A recent technique \cite{Bewley2006a} allows us to visualize the dynamics of quantized vortices in superfluid $^4$He using micron-sized frozen particles, created by injecting a gaseous helium-hydrogen mixture above the lambda transition. The tracers, trapped on the vortex cores because of a Bernoulli pressure gradient, are illuminated with a laser-sheet and imaged with a CCD camera.
In particular, this technique allows us to directly characterize quantized vortex reconnection \cite{Sreenivasan2008}, and to clearly distinguish classical and quantum turbulence \cite{PaolettiPRL2008}.

Using a new technique to create sub-micron sized frozen particles directly in the superfluid state, we directly visualize for the first time the helical waves excited after a reconnection event.
These waves, named after the pioneering work of Kelvin \cite{Thomson1880}, have been studied in different flow phenomena like trailing wakes and tornadoes, and are understood to be involved in quantum turbulence dissipation in the vanishing viscosity, zero-temperature limit. \cite{Svistunov1995}.

\begin{figure}
\centerline{\includegraphics[width=.8\textwidth]{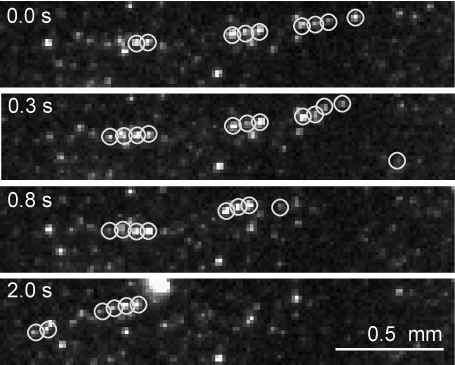}}
\caption{Sequence of images of a wave event captured at 54 frames per second at a temperature of 1.98K}\label{fig:wave}
\end{figure}

Figure~\ref{fig:wave} shows a sequence of images of one of these waves captured at 54 frames per second at a temperature of 1.98K, where the circles highlight the particles trapped on the vortex line. 
The evolution of the vortex shape compares well with the self-similar solutions of vortex filaments models. In particular we compare the data with the local induction approximation (LIA) \cite{Schwarz1985} as well as the fully non-local Biot--Savart equations \cite{Hormoz2012} with a phenomenological damping term.
More details are available in \cite{Fondac}.

\end{document}